\documentclass[twoside,12pt]{article}
\usepackage{CJK}
\usepackage{indentfirst}
\usepackage{bm}
\usepackage{graphicx}

\begin{document}

\begin{CJK*}{GBK}{song}

\begin{center}
\LARGE\bf Construction of Three-Qubit Kochen-Specker Sets$^{}$
\end{center}

\footnotetext{\hspace*{-.45cm}\footnotesize $^\dag$Corresponding author. E-mail: SingPoh.Toh@nottingham.edu.my; singpoh@gmail.com }

\begin{center}
\rm S.P. Toh $^\dagger$
\end{center}

\begin{center}
\begin{footnotesize} \sl
Faculty of Engineering, The University of Nottingham Malaysia Campus \\
Jalan Broga, 43500 Semenyih, Selangor Darul Ehsan, Malaysia \\
\end{footnotesize}
\end{center}

\vspace*{2mm}

\begin{center}
\begin{minipage}{15.5cm}
\parindent 20pt\footnotesize
Contextuality is one of the fundamental deviations of quantum mechanics from classical physics. The Kochen-Specker (KS) theorem shows that non-contextual classical physics with hidden variables is inconsistent with the predictions of quantum mechanics. Parity proof is one of the many different approaches applied to prove KS theorem for quantum system with different dimensionality. This method of proof requires KS sets that composed of an odd number of bases and an even number of projectors. In most of the cases, the number of KS sets produced is large and its production is generally aided by computer calculation. However, manually generation of  KS sets are also reported previously for qutrit and two-qubit systems. We put forward the first and surprisingly simple method to generate manually KS sets, with respect to Mermin pentagram, that can be used for the parity proof of KS theorem in a state space of eight-dimensional three-qubit system .
\end{minipage}
\end{center}

\begin{center}
\begin{minipage}{15.5cm}
\begin{minipage}[t]{2.3cm}{\bf Keywords:}
\end{minipage}
\begin{minipage}[t]{13.1cm}
Kochen-Specker theorem, contextuality, hidden variable, three-qubit
\end{minipage}\par\vglue8pt
{\bf PACS: }
03.65.Aa, 03.65.Ta, 42.50.Dv
\end{minipage}
\end{center}

\newpage

\section{Introduction}
The Kochen-Specker theorem ${[1]}$ states that in a Hilbert space with a finite dimension, $d \geq 3$, it is possible to produce a set of $n$ projectors, representing yes-no questions about a quantum system, such that none of the $2^n$ possible answer is compatible with the sum rules of quantum mechanics for orthogonal resolution of the identity. The physical meaning of the Kochen-Specker (KS) theorem is that quantum mechanics is contextual and no way to be simulated by any noncontextual hidden-variable (NCHV) theories. Quantum contextuality is a statement that the result of an experiment in a hidden variable theory must depend on which other compatible experiments are performed alongside, if the hidden variable theory is to be compatible with quantum theory. Two observables $A$ and $B$ are mutually compatible if the result of $A$ does not depend whether $B$ is measured before, after or simultaneously with $A$ and vice versa. NCHV theories assume that result of a measurement of an observable is predetermined and independent of measurement context. A set of maximally collection of compatible observable defines a context.

\vspace*{5mm}

The first theoretical proof of KS theorem is reported by Kochen and Specker that used 117 projective measurements in the real three-dimensional Hilbert space ${[1]}$.  Peres' proof ${[2]}$ drastically reduces the number of necessary measurements to 33 and 24 projectors (or rays) for three- and four-dimensional systems, respectively. Up to now the smallest numbers of rays required in the proof of KS theorem are 31 ${[3]}$,  18 ${[4]}$
and 36 ${[5]}$ in three-, four- and eight-dimensional systems, respectively. Operators that represent observables also used to prove KS theorem. Mermin ${[6]}$ used an array of nine observables for two spin-$\frac{1}{2}$ particles to show some kinds of logical contradictions that lead to KS proof. Similar mathematical simplicity is also demonstated in KS proof for a three-qubit system using ten observables ${[6]}$.

\vspace*{5mm}

The doubt about the possibility of testing KS theorem experimentally in the real world was once arose because of the unavoidable finite measurement times and precision ${[7,8]}$.  Nowadays, KS theorem becomes experimentally testable due to KS inequalities ${[9]}$ that obeyed by all the NCHV models but violated by quantum mechanics. Recent affirmative results were reported for experiments performed on different physics systems such as a pair of trapped ions ${[10]}$, neutrons ${[11]}$, single photons ${[12]}$, two photonic qubits ${[13]}$ and nuclear spins ${[14]}$.

\vspace*{5mm}

One of the typical methods to prove KS theorem relies on the search of KS set. A KS set in dimensional $d$ is defined as a set $S$ of $d$-dimensional complex vectors, with $d \geq 3$ and with the property that there is no map $f:S \to \{0,1\}$ such that, for any context that represented by an orthogonal basis in $S$, one and only one of the vectors is mapped to 1 ${[1]}$. Since the number of KS sets are in generally huge, many constructions are obtained via computer calculation ${[15,16]}$. However, there are few papers discuss elegantly the whole construction process of KS sets, for example Bub ${[17]}$ showed the details of generating 33 rays based on Sch$\ddot{\mbox{u}}$tte's tautology and Waegell and Aravind showed the ways of constructing KS sets from 24 rays of Peres ${[18]}$ and also that based on 60 complex rays ${[19]}$  in four dimensions. We will present explicitly a simple method to manually construct KS sets for three-qubit system with respect to the Mermin pentagram.

\vspace*{5mm}

The plan of this paper is as follows. In Section 2, we briefly introduce the important features for the Mermin pentagram and give the 40 common eigenvectors that derived from the mutually commuting sets of observables in Mermin pentagram. We then explain the physical meaning of KS theorem by presenting a parity proof that using 36 rays and 11 bases for three-qubit system. The method of constructing this type of KS set as well as the other two types with 38 rays and 40 rays, respectively, will be presented in Section 3. Finally, in Section 4, we conclude by relating the patterns of constructions between our method and previous works.

\section{Kernaghan and Peres' 40 rays}
Mermin used 10 observables, which are arranged in five groups of four, for a three qubit system to form a pentagram as shown in Figure 1 ${[6]}$. Each observable has only the eigenvalues 1 or -1. The observables in any line of the pentagram form a mutually commuting set. The product of the observables in any line of the pentagram is +1, with exception of the observabes on the horizontal line for which this product is -1. This means that there are 5 measurement contexts depicted in Figure 1. Although a KS proof with the 10 observables that each repeat twice in the 5 measurement contexts can be shown ${[6]}$, the particular interest for this work is KS proof using projection operators (or projectors), as explained in the following.  The common eigenvectors (or rays) for the five sets of four mutually commuting operators are shown in Table 1 ${[5]}$. Note that the 40 rays are labeled as  $R_i$ with \mbox{$i=$1, 2, 3, \textellipsis, 40} and the symbol $\bar{1}$ is used to denote $-1$.

\vspace*{10mm}

\centerline{\includegraphics{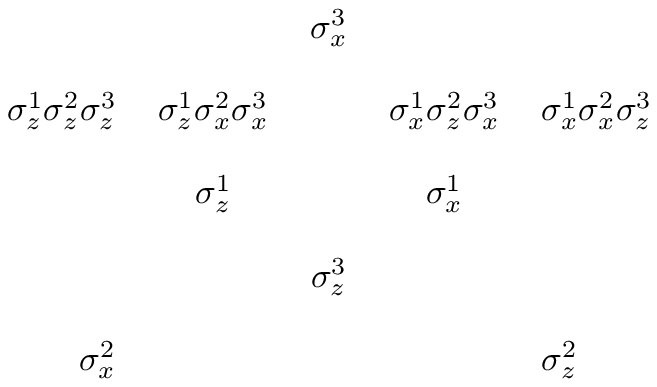}}

\begin{center}
\parbox{15.5cm}{\small{\bf Fig.1} Mermin Pentagram. The three-qubit observables  are arranged at the vertices of a pentagram. The four observables along any line are pairwise commuting. Note that $\sigma_x$, $\sigma_y$ and $\sigma_z$ are operators of a qubit. The subscripts 1, 2 and 3 on the operators referring to the first, second and third qubit, respectively. The product of the four observables on every line of the pentagram but the horizontal line is 1. The product of the four observables on the horizontal line is -1. }
\end{center}

\begin{table}[!h]
\caption{The common eigenvectors for \{$\sigma^1_z\sigma^2_z\sigma^3_z$, $\sigma^1_z$, $\sigma^2_z$, $\sigma^3_z$\}, \{$\sigma^1_z\sigma^2_x\sigma^3_x$, $\sigma^1_z$, $\sigma^2_x$, $\sigma^3_x$\}, \{$\sigma^1_x\sigma^2_z\sigma^3_x$, $\sigma^1_x$, $\sigma^2_z$, $\sigma^3_x$\}, \{$\sigma^1_x\sigma^2_x\sigma^3_z$, $\sigma^1_x$, $\sigma^2_x$, $\sigma^3_z$\} and \{$\sigma^1_z\sigma^2_x\sigma^3_x$, $\sigma^1_x\sigma^2_z\sigma^3_x$, $\sigma^1_x\sigma^2_x\sigma^3_z$, $\sigma^1_z\sigma^2_z\sigma^3_z$\} listed from the left to the right column. These 40 rays, labeled from $R_1$ to $R_{40}$, were derived by Kernaghan and Peres from the Mermin pentagram. The symbol $\bar{1}$ is used to denote $-1$.}
\begin{center}
\begin{tabular}{|c|c|c|c|c|c|c|c|c|c|}
  \hline
  $R_1$ & 10000000 & $R_9$ & 11110000 & $R_{17}$ & 11001100 & $R_{25}$ & 10101010 & $R_{33}$ & 100101$\bar{1}$0 \\
  $R_2$ & 01000000 & $R_{10}$ & 11$\bar{1}$$\bar{1}$0000 & $R_{18}$ & 1100$\bar{1}$$\bar{1}$00 & $R_{26}$ & 1010$\bar{1}$0$\bar{1}$0 & $R_{34}$ & 100$\bar{1}$0110 \\
  $R_3$ & 00100000 & $R_{11}$ & 1$\bar{1}$$1\bar{1}$0000 & $R_{19}$ & 1$\bar{1}$001$\bar{1}$00 & $R_{27}$ & 10$\bar{1}$010$\bar{1}$0 & $R_{35}$ & 10010$\bar{1}$10 \\
  $R_4$ & 00010000 & $R_{12}$ & 1$\bar{1}$$\bar{1}$10000 & $R_{20}$ & 1$\bar{1}$00$\bar{1}$100 & $R_{28}$ & 10$\bar{1}$0$\bar{1}$010 & $R_{36}$ & 100$\bar{1}$0$\bar{1}$$\bar{1}$0 \\
  $R_5$ & 00001000 & $R_{13}$ & 00001111 & $R_{21}$ & 00110011 & $R_{29}$ & 01010101 & $R_{37}$ & 0110$\bar{1}$001 \\
  $R_6$ & 00000100 & $R_{14}$ & 000011$\bar{1}$$\bar{1}$ & $R_{22}$ & 001100$\bar{1}$$\bar{1}$ & $R_{30}$ & 01010$\bar{1}$$0\bar{1}$ & $R_{38}$ & 01$\bar{1}$01001 \\
  $R_7$ & 00000010 & $R_{15}$ & 00001$\bar{1}$1$\bar{1}$ & $R_{23}$ & 001$\bar{1}$001$\bar{1}$ & $R_{31}$ & 010$\bar{1}$010$\bar{1}$ & $R_{39}$ & 0$\bar{1}$101001 \\
  $R_8$ & 00000001 & $R_{16}$ & 00001$\bar{1}$$\bar{1}$1 & $R_{24}$ & 001$\bar{1}$00$\bar{1}$1 & $R_{32}$ & 010$\bar{1}$0$\bar{1}$01 & $R_{40}$ & 0$\bar{1}$$\bar{1}$0$\bar{1}$001 \\
  \hline
\end{tabular}
\end{center}
\end{table}

The outer product of a ray will give rise to a rank-1 projector, which will be denoted as $P_i$, with \mbox{$i=$1, 2, 3, \textellipsis, 40}. Based on these 40 projectors,  Kernaghan and Peres constructed a KS set with 36 rays and 11 bases, as shown below, to prove KS theorem ${[5]}$.

\begin{eqnarray} \label{eqA}
\hat{P}_{1}+\hat{P}_{2}+\hat{P}_{3}+\hat{P}_{4}+\hat{P}_{5}+\hat{P}_{6}+\hat{P}_{7}+\hat{P}_{8} &=& I, \nonumber \\
\hat{P}_{1}+\hat{P}_{2}+\hat{P}_{3}+\hat{P}_{4}+\hat{P}_{13}+\hat{P}_{14}+\hat{P}_{15}+\hat{P}_{16} &=& I, \nonumber \\
\hat{P}_{1}+\hat{P}_{2}+\hat{P}_{5}+\hat{P}_{6}+\hat{P}_{21}+\hat{P}_{22}+\hat{P}_{23}+\hat{P}_{24} &=& I, \nonumber \\
\hat{P}_{1}+\hat{P}_{3}+\hat{P}_{5}+\hat{P}_{7}+\hat{P}_{29}+\hat{P}_{30}+\hat{P}_{31}+\hat{P}_{32} &=& I, \nonumber \\
\hat{P}_{2}+\hat{P}_{3}+\hat{P}_{5}+\hat{P}_{8}+\hat{P}_{33}+\hat{P}_{34}+\hat{P}_{35}+\hat{P}_{36} &=& I, \nonumber \\
\hat{P}_{9}+\hat{P}_{10}+\hat{P}_{13}+\hat{P}_{14}+\hat{P}_{19}+\hat{P}_{20}+\hat{P}_{23}+\hat{P}_{24} &=& I, \nonumber \\
\hat{P}_{9}+\hat{P}_{11}+\hat{P}_{13}+\hat{P}_{15}+\hat{P}_{27}+\hat{P}_{28}+\hat{P}_{31}+\hat{P}_{32} &=& I, \nonumber \\
\hat{P}_{10}+\hat{P}_{11}+\hat{P}_{13}+\hat{P}_{16}+\hat{P}_{33}+\hat{P}_{35}+\hat{P}_{37}+\hat{P}_{40} &=& I, \nonumber \\
\hat{P}_{17}+\hat{P}_{19}+\hat{P}_{21}+\hat{P}_{23}+\hat{P}_{26}+\hat{P}_{28}+\hat{P}_{30}+\hat{P}_{32} &=& I, \nonumber \\
\hat{P}_{17}+\hat{P}_{20}+\hat{P}_{22}+\hat{P}_{23}+\hat{P}_{35}+\hat{P}_{36}+\hat{P}_{37}+\hat{P}_{39} &=& I, \nonumber \\
\hat{P}_{26}+\hat{P}_{27}+\hat{P}_{29}+\hat{P}_{32}+\hat{P}_{34}+\hat{P}_{35}+\hat{P}_{39}+\hat{P}_{40} &=& I.
\end{eqnarray}

\vspace*{5mm}

The eigenvalues of the above 36 rank-1 projectors are 0 or 1 and the eigenvalue of $8 \times 8$ identity matrix, $I$ is 1. Since each of the projectors on the left of equality signs repeat either twice or four times , the summation of the values assigned to all of them must result in an even number. On the other hand, the summation of all the values assigned to the identity matrices on the right of equality signs is 11, which is an odd number, a contradiction thus occurs. In the process of value assignment, the values are presumed to be determined by hidden variables and pre-exists before measurement. In addition, the value assignment is done based on the non-contextuality assumption in which all the repeated projectors must be assigned the same value independent of the bases or measurement contexts. The above contradiction can be removed if the values pre-defined to the same projectors are allowed to alter when the measurement contexts change.  Thus the contradiction demonstrated in KS proof shows that even if there are hidden variables that predetermined the values of measurements, it must depend on the measurement contexts.  This is what meant by stating that non-contextual hidden variable theories are incompatible to contextual quantum mechanics.

\vspace*{5mm}

The method of KS proof that showing the impossibility of consistent value assignment between an even number of rays and an odd number of bases is called parity proof. Parity proofs are of interest because they can be used to derive KS inequalities for testing quantum contextuality ${[15]}$. The proof shown above is an example of parity proof and the rays used constitute a KS set. In the above KS set, there are 28 rays (or projectors, since an outer product of a ray produces a projector) that each repeat twice and 8 rays that each occur 4 times. The total 36 rays form 11 bases with 8 rays in a basis. All these information is given in the symbol $28_{2}8_{4}\mbox{-}11_{8}$. Some years later,  Waegell and Aravind ${[16]}$ found, through computer search, KS sets with the form of $24_{2}14_{4}\mbox{-}13_{8}$ and $20_{2}20_{4}\mbox{-}15_{8}$, and thus accomplished the complete search of all 3-qubit KS sets with respect to Mermin pentagram. With the above background, we are now ready to move on to the next section to put forward the method of constructing all these KS sets manually.

\section{Three Ways to Constsruct KS sets}
All the KS sets that generated from the 40-ray of Kernaghan-Peres set can be easily constructed by the steps proposed in this section. Before the full description about the process, some preparations are required. Based on the 40 rays in Table 1, 25 bases are constructed with 8 rays in a basis. We list all these bases in the circles shown in Figure 2 ${[16]}$. The first 5 bases composed of rays shown in the five columns of Table 1 are represented in the 5 inner circles. These 5 bases are called pure bases, which will be denoted as $x_i$ with \mbox{$i=$1, \textellipsis, 5}. Any two pure bases will give rise to two hybrid bases and the corresponding four bases are joint by a straight line. There are in total 20 hybrid bases which are denoted as $y_i$ with \mbox{$i=$1, \textellipsis, 20} and the numbering is shown beside the outer circles in Figure 2. Two hybrid bases that are connected by a straight line are labeled as $y\mbox{-}y'$. As an example, the pure bases $x_1$ and $x_3$ produce $y_7$ and $y_{12}$, therefore $x_1$, $x_3$, $y_7$ and $y_{12}$ are connected by a straight line.

\centerline{\includegraphics{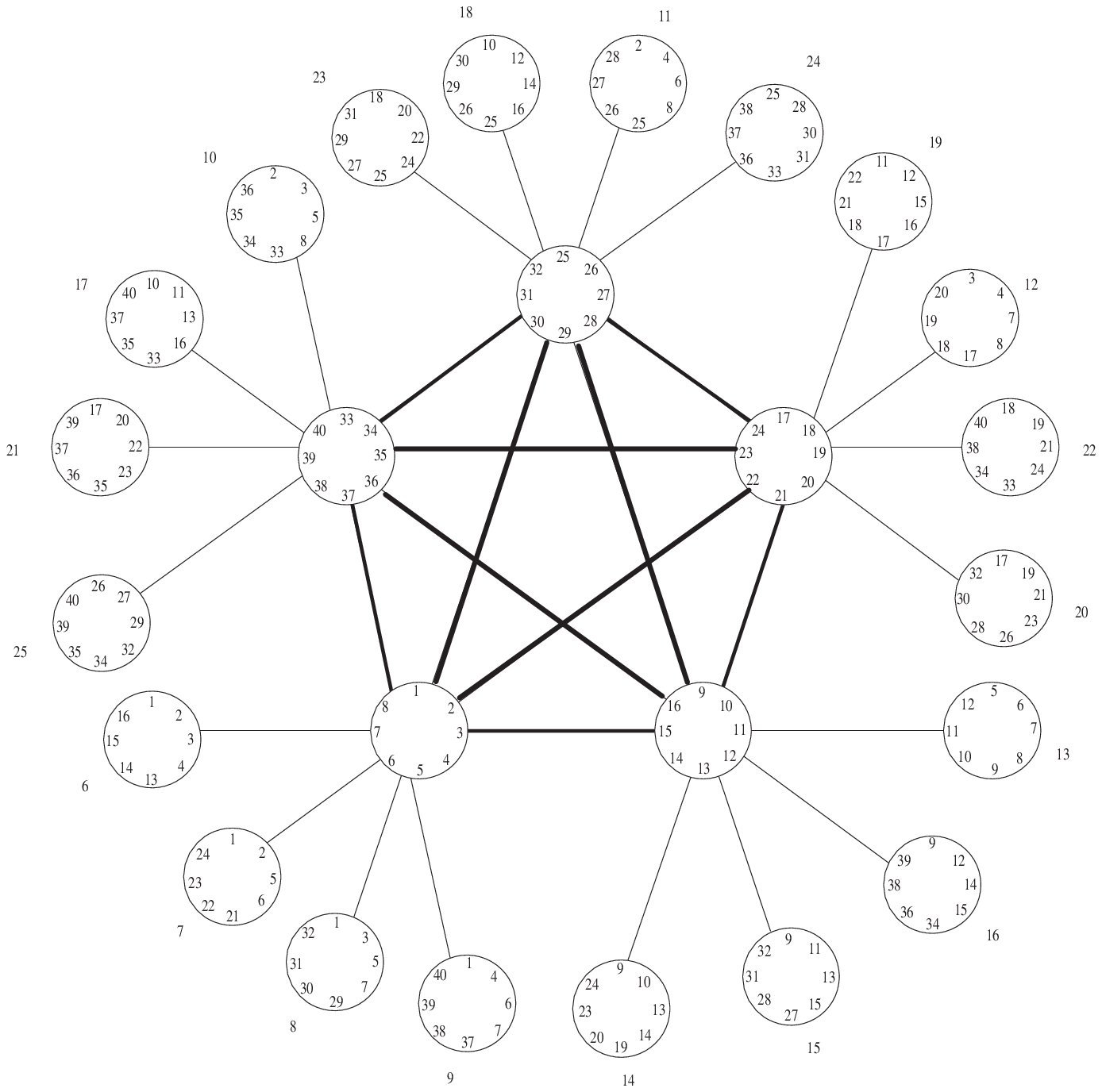}}

\begin{center}
\parbox{15.5cm}{\small{\bf Fig.2} The 25 bases derived from the 40 rays of Kernaghan and Peres. Each circle represents a basis with the 8 rays listed inside. The inner circles represent pure bases and outer circles represent hybrid bases. The hybrid bases joint by a straight line are obtained equally from two pure bases that lie on the line. Numbering for the 20 hybrid bases are shown beside the circles.}
\end{center}

\vspace*{5mm}

The key of our method rests on the selection of sets of 4 rays that  listed in Table 2. These sets of 4 rays are labeled as $\Gamma^k_{ij}$, where \mbox{$i=$1, \textellipsis, 5} and \mbox{$j=$1, \textellipsis, 8} that refer to the columns and rows of Table 2, respectively, and the subscript $k$ refers to the $k$-th ($k=1, 2, 3$) of $\Gamma_{ij}$ chosen. We may also interpret \mbox{$i=$1, \textellipsis, 5} as referring to the pure bases where these sets of 4 rays derived from. Besides, in our method of construction, it also requires set of 3 elements, denoted as $L_k$, that taken from $\Gamma^k_{ij}$. As an example, if the first set of 4 rays chosen is $\Gamma^1_{11}=\{1,2,3,5\}$, then $L_1=\{\{1,2,3\},\{1,2,5\},\{1,3,5\},\{2,3,5\}\}$.

\begin{table*}[htb]
\caption{Sets of 4 rays extracted from the 5 pure bases. Each time a set of 4 rays, denoted as $\Gamma^k_{ij}$, be selected, set of 3 elements, denoted as $L_k$ is formed,  and both $\Gamma^k_{ij}$ and $L_k$ guide the selection of bases required to prove KS theorem. }
\begin{center}
\begin{tabular}{|c|c|c|c|c|c|}
  \hline
  $j$ & $\Gamma^k_{1j}$ & $\Gamma^k_{2j}$ & $\Gamma^k_{3j}$ & $\Gamma^k_{4j}$ & $\Gamma^k_{5j}$ \\ \hline
  1 & 1\hspace{2mm}2\hspace{2mm}3\hspace{2mm}5  & \hspace{2mm}9\hspace{2mm}10\hspace{2mm}11\hspace{2mm}13  & 17\hspace{2mm}18\hspace{2mm}19\hspace{2mm}21 & 25\hspace{2mm}26\hspace{2mm}27\hspace{2mm}29 & 33\hspace{2mm}34\hspace{2mm}35\hspace{2mm}40 \\
  2 & 1\hspace{2mm}2\hspace{2mm}4\hspace{2mm}6  & \hspace{2mm}9\hspace{2mm}10\hspace{2mm}12\hspace{2mm}14  & 17\hspace{2mm}18\hspace{2mm}20\hspace{2mm}22 & 25\hspace{2mm}26\hspace{2mm}28\hspace{2mm}30 & 33\hspace{2mm}34\hspace{2mm}36\hspace{2mm}38 \\
  3 & 1\hspace{2mm}3\hspace{2mm}4\hspace{2mm}7  & \hspace{2mm}9\hspace{2mm}11\hspace{2mm}12\hspace{2mm}15  & 17\hspace{2mm}19\hspace{2mm}20\hspace{2mm}23 & 25\hspace{2mm}27\hspace{2mm}28\hspace{2mm}31 & 33\hspace{2mm}35\hspace{2mm}36\hspace{2mm}37 \\
  4 & 1\hspace{2mm}5\hspace{2mm}6\hspace{2mm}7  & \hspace{2mm}9\hspace{2mm}13\hspace{2mm}14\hspace{2mm}15  & 17\hspace{2mm}21\hspace{2mm}22\hspace{2mm}23 & 25\hspace{2mm}29\hspace{2mm}30\hspace{2mm}31 & 33\hspace{2mm}37\hspace{2mm}38\hspace{2mm}40 \\
  5 & 2\hspace{2mm}3\hspace{2mm}4\hspace{2mm}8  & 10\hspace{2mm}11\hspace{2mm}12\hspace{2mm}16 & 18\hspace{2mm}19\hspace{2mm}20\hspace{2mm}24 & 26\hspace{2mm}27\hspace{2mm}28\hspace{2mm}32 & 34\hspace{2mm}35\hspace{2mm}36\hspace{2mm}39 \\
  6 & 2\hspace{2mm}5\hspace{2mm}6\hspace{2mm}8  & 10\hspace{2mm}13\hspace{2mm}14\hspace{2mm}16 & 18\hspace{2mm}21\hspace{2mm}22\hspace{2mm}24 & 26\hspace{2mm}29\hspace{2mm}30\hspace{2mm}32 & 34\hspace{2mm}38\hspace{2mm}39\hspace{2mm}40 \\
  7 & 3\hspace{2mm}5\hspace{2mm}7\hspace{2mm}8  & 11\hspace{2mm}13\hspace{2mm}15\hspace{2mm}16 & 19\hspace{2mm}21\hspace{2mm}23\hspace{2mm}24 & 27\hspace{2mm}29\hspace{2mm}31\hspace{2mm}32 & 35\hspace{2mm}37\hspace{2mm}39\hspace{2mm}40 \\
  8 & 4\hspace{2mm}6\hspace{2mm}7\hspace{2mm}8  & 12\hspace{2mm}14\hspace{2mm}15\hspace{2mm}16 & 20\hspace{2mm}22\hspace{2mm}23\hspace{2mm}24 & 28\hspace{2mm}30\hspace{2mm}31\hspace{2mm}32 & 36\hspace{2mm}37\hspace{2mm}38\hspace{2mm}39 \\
  \hline
\end{tabular}
\end{center}
\end{table*}

We now present detailed steps of construction for the above-mentioned three types of KS sets with respect to Mermin pentagram. The following are steps to construct $28_{2}8_{4}\mbox{-}11_{8}$ KS sets:

\vspace*{5mm}

\indent \indent $S1$ : Choose $\Gamma^1_{ij}$. \\
\indent \indent $S2$ : Tick off $x$ that contains $\Gamma^1_{ij}$. \\
\indent \indent $S3$ : Tick off those $y$ that contain $L_1$ and cross out the corresponding $y'$. Let these \\
\indent \hspace{1.5cm} ticked off $y$ form set $J$. \\
\indent \indent $S4$ : Pick a new ray (i.e., ray not contained in $L_1$) in $J$, examine if it existed in the \\
\indent \hspace{1.5cm} crossed out $y'$. If it is, choose the next ray for examination. \enspace Otherwise, tick \\
\indent \hspace{1.5cm} off all $y$ containing that ray, and again cross out the corresponding $y'$. \\
\indent \indent $S5$ : Repeat $S4$ for other rays in $J$ until all $y$ are checked off \enspace (either ticked off \\
\indent \hspace{1.5cm} or crossed out). \\

After the execution of the above 5 steps, all the ticked off $y$ and the single ticked off $x$ form a $28_{2}8_{4}\mbox{-}11_{8}$ KS set. For the same chosen $\Gamma^1_{ij}$, the sequence of the new rays picked for examination in $S4$ and $S5$ will produce different $28_{2}8_{4}\mbox{-}11_{8}$ KS sets. As an example, if $\Gamma^1_{11}$ was chosen in $S1$, $R_{13}$ was chosen in $S4$ followed by $R_{23}$ and $R_{32}$ in $S5$ for examination, the KS set discussed in Section 2 would be obtained. Figure 3 displays how the completed work looks like after carrying out the above construction process.

\centerline{\includegraphics{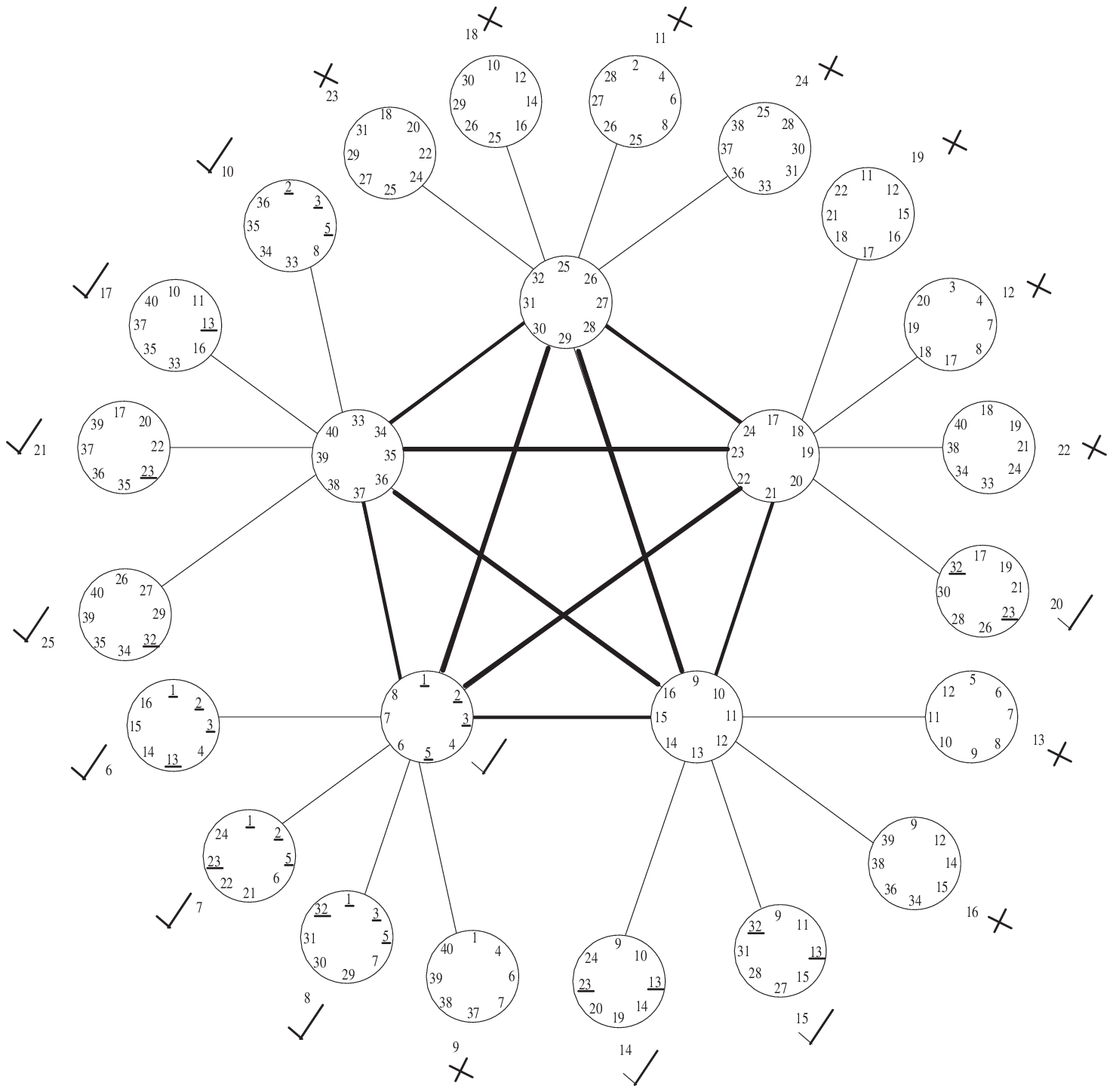}}

\begin{center}
\parbox{15.5cm}{\small{\bf Fig.3} Completed diagram after the execution of $S1\mbox{-}S5$ for a construction of $28_{2}8_{4}\mbox{-}11_{8}$ KS set. In this example, $\Gamma^1_{11}$ was first chosen, followed by $R_{13}$, $R_{23}$ and $R_{32}$. The selected rays are underlined. Ticked off bases form the KS set with 11 bases or measurement contexts that given explicitly in (1).}
\end{center}

\vspace*{5mm}

Only slight modification of the above procedure is needed to construct $24_{2}14_{4}\mbox{-}13_{8}$ KS sets. The complete steps are described as below:

\vspace*{5mm}

\indent \indent $S1$ : Choose $\Gamma^1_{ij}$ \\
\indent \indent $S2$ : Tick off $x$ that contains $\Gamma^1_{ij}$. \\
\indent \indent $S3$ : Tick off those $y$ that contain $L_1$ and cross out the corresponding $y'$. \\
\indent \indent $S4$ : Repeat $S1$-$S3$ for $\Gamma^2_{ij}$ \\
\indent \indent $S5$ : Repeat $S1$-$S3$ for $\Gamma^3_{ij}$ \\
\indent \indent $S6$ : After $S5$, it remains one $y\mbox{-}y'$ which has not been checked off yet. \enspace Tick off that  \\
\indent \hspace{1.5cm} basis which containing 2 rays that each repeated 3 times in the previously ticked \\
\indent \hspace{1.5cm} off $y$.

\vspace*{5mm}

Note that $\Gamma^1_{ij}$, $\Gamma^2_{ij}$ and $\Gamma^3_{ij}$ should not be picked from the same column of Table 2. The completion of the above steps gives rise to 3 pure bases and 10 hybrid bases that form a $24_{2}14_{4}\mbox{-}13_{8}$ KS sets. Note that while carrying out $S1$-$S6$, any step producing crossed out $y\mbox{-}y'$ will incur the failure of the construction.

\vspace*{5mm}

The construction of $20_{2}20_{4}\mbox{-}15_{8}$ KS sets is the simplest, as follow:

\vspace*{5mm}

\indent \indent $S1$ : Choose $\Gamma^1_{ij}$ \\
\indent \indent $S2$ : Tick off $x$ that contains $\Gamma^1_{ij}$. \\
\indent \indent $S3$ : Tick off those $y$ that contain $L_1$ and cross out the corresponding $y'$. \\
\indent \indent $S4$ : Repeat $S1$-$S3$ for $\Gamma^2_{ij}$ \\
\indent \indent $S5$ : Repeat $S1$-$S3$ for $\Gamma^3_{ij}$ \\
\indent \indent $S6$ : Repeat $S1$-$S3$ for $\Gamma^4_{ij}$ \\

Again, note that $\Gamma^1_{ij}$, $\Gamma^2_{ij}$, $\Gamma^3_{ij}$ and $\Gamma^4_{ij}$ should not be picked from the same column of Table 2. Any step producing crossed out $y\mbox{-}y'$ will again incur the failure of the construction.

\vspace*{5mm}

The above three ways of constructing KS sets from 40-ray Kernaghan-Peres set clearly show the key role played by the sets of 4 rays given in Table 2. Only 1, 3 and 4 $\Gamma^k_{ij}$ are needed to construct KS sets with 11, 13 and 15 bases, respectively.

\section{Conclusion}
As one of the typical methods of Kochen-Specker (KS) theorem proof, parity proof that relies on KS sets was frequently reported. The number of KS sets is generally huge and most of them are produced through exhaustive computer search. Nonetheless, there are no lack of manual construction, for example those that reported by Bub ${[17]}$ for qutrit system and by Waegell and Aravind ${[18,19]}$ for two-qubit system. We extend the list by putting forward a simple method to construct manually three different types of KS sets for a three-qubit quantum system with respect to the Mermin pentagram. These KS sets are recently used to experimentally test the quantum contextuality ${[20]}$. Our method is simple in the sense that it involves merely the selection of 4-ray sets and few steps of examination. The simplicity of our method is arisen from the relationships between pure and hybrid bases depicted in Figure 2. The similar relationships can also be found for two- qubit system ${[18,19]}$.  It is thus interesting to investigate if these relationships also exist in higher dimensional quantum system. More specifically, it is worthy to focus the next investigation on finding out if other ${[16]}$ observable-based KS proofs for three-qubit system can be converted into projector-based proofs by adopting the similar construction pattern for KS sets.

\vspace*{5mm}

\noindent {\bf List of Captions}

\noindent Figure 1: Mermin Pentagram. The three-qubit observables  are arranged at the vertices of a pentagram. The four observables along any line are pairwise commuting. Note that $\sigma_x$, $\sigma_y$ and $\sigma_z$ are operators of a qubit. The subscripts 1, 2 and 3 on the operators referring to the first, second and third qubit, respectively. The product of the four observables on every line of the pentagram but the horizontal line is 1. The product of the four observables on the horizontal line is -1. \\

\noindent Figure 2: The 25 bases derived from the 40 rays of Kernaghan and Peres. Each circle represents a basis with the 8 rays listed inside. The inner circles represent pure bases and outer circles represent hybrid bases. The hybrid bases joint by a straight line are obtained equally from two pure bases that lie on the line. Numbering for the 20 hybrid bases are shown beside the circles.\\

\noindent Figure 3: Completed diagram after the execution of $S1\mbox{-}S5$ for a construction of $28_{2}8_{4}\mbox{-}11_{8}$ KS set. In this example, $\Gamma^1_{11}$ was first chosen, followed by $R_{13}$, $R_{23}$ and $R_{32}$. The selected rays are underlined. Ticked off bases form the KS set with 11 bases or measurement contexts that given explicitly in (1).\\

\noindent Table 1: The common eigenvectors for \{$\sigma^1_z\sigma^2_z\sigma^3_z$, $\sigma^1_z$, $\sigma^2_z$, $\sigma^3_z$\}, \{$\sigma^1_z\sigma^2_x\sigma^3_x$, $\sigma^1_z$, $\sigma^2_x$, $\sigma^3_x$\}, \{$\sigma^1_x\sigma^2_z\sigma^3_x$, $\sigma^1_x$, $\sigma^2_z$, $\sigma^3_x$\}, \{$\sigma^1_x\sigma^2_x\sigma^3_z$, $\sigma^1_x$, $\sigma^2_x$, $\sigma^3_z$\} and \{$\sigma^1_z\sigma^2_x\sigma^3_x$, $\sigma^1_x\sigma^2_z\sigma^3_x$, $\sigma^1_x\sigma^2_x\sigma^3_z$, $\sigma^1_z\sigma^2_z\sigma^3_z$\} listed from the left to the right column. These 40 rays, labeled from $R_1$ to $R_{40}$, were derived by Kernaghan and Peres from the Mermin pentagram. The symbol $\bar{1}$ is used to denote $-1$.\\

\noindent Table 2: Sets of 4 rays extracted from the 5 pure bases. Each time a set of 4 rays, denoted as $\Gamma^k_{ij}$, be selected, set of 3 elements, denoted as $L_k$ is formed,  and both $\Gamma^k_{ij}$ and $L_k$ guide the selection of bases required to prove KS theorem.

\end{CJK*}

\begin{thebibliography}{99}
\itemsep=-4pt plus.2pt minus.2pt
\small
\bibitem{R1} K. Kochen, E. P. Specker, {\it J. Math. Mech.} {\bf 17} 59 (1967). DOI:dx.doi.org/10.1512/iumj.1968.17.17004
\bibitem{R2} A. Peres, {\it J. Phys. A: Math. Gen.} {\bf 24} L175 (1991). DOI:10.1088/0305-4470/24/4/003
\bibitem{R3} J. H. Conway,  S. Kochen, reported by A. Peres, {\it Quantum Theory: Concepts and Method} Dordrecht: Kluwer 1993, p.114.
\bibitem{R4} A. Cabello, J. M. Estebaranz, G. Garc\'{i}a-Alcaine, {\it Phys. Lett.} A {\bf 212} 183 (1996). DOI:10.1016/0375-9601(96)00134-X
\bibitem{R5} M. Kernaghan, A. Peres, {\it Phys. Lett.,} A {\bf 198} 1 (1995). DOI: 10.1016/0375-9601(95)00012-R
\bibitem{R6} N. D. Mermin, {\it Rev. Mod. Phys.} {\bf 65} 803 (1993). DOI: http://dx.doi.org/10.1103/RevModPhys.65.803
\bibitem{R7} D. A. Meyer,  {\it Phys. Rev. Lett.} {\bf 83} 3751 (1999). DOI: http://dx.doi.org/10.1103/PhysRevLett.83.3751
\bibitem{R8} A. Kent, {\it Phys. Rev. Lett.} {\bf 83}  3755 (1999). DOI: http://dx.doi.org/10.1103/PhysRevLett.83.3755
\bibitem{R9} A. Cabello, {\it Phys. Rev. Lett.} {\bf 101} 210401 (2008). DOI: http://dx.doi.org/10.1103/PhysRevLett.101.210401
\bibitem{R10} G. Kirchmair, F. Z$\ddot{\mbox{a}}$hringer, R. Gerritsma, M. Kleinmann, O. G$\ddot{\mbox{u}}$hne, A. Cabello, R. Blatt, C. F. Roos, {\it Nature} {\bf 460} 494 (2009). DOI: 10.1038/nature08172
\bibitem{R11} H. Bartosik, J. Klepp, C. Schmitzer, S. Sponar, A. Cabello, H. Rauch, Y. Hasegawa, {\it Phys. Rev. Lett.} {\bf 103} 040403 (2009). DOI: http://dx.doi.org/10.1103/PhysRevLett.103.040403
\bibitem{R12} E. Amselem, M. Radmark, M. Bourennane, A. Cabello, {\it Phys. Rev. Lett.} {\bf 103} 160405 (2009). DOI: http://dx.doi.org/10.1103/PhysRevLett.103.160405
\bibitem{R13} G. Borges, M. Carvalho, P. L. de Assis, J. Ferraz, M. Ara$\acute{\mbox{u}}$jo, A. Cabello, M. Terra Cunha, S. P$\acute{\mbox{a}}$dua, {\it Phys. Rev.} A {\bf 89} 052106 (2014). DOI: http://dx.doi.org/10.1103/PhysRevA.89.052106
\bibitem{R14} O. Moussa, C. A. Ryan, D. G. Cory, R. Laflamme, {\it Phys. Rev. Lett.} {\bf 104} 160501 (2010). DOI: http://dx.doi.org/10.1103/PhysRevLett.104.160501
\bibitem{R15} M. Waegell, P. K. Aravind, {\it Found. Phys.} {\bf 44} 1085 (2014). DOI: 10.1007/s10701-014-9830-0
\bibitem{R16} M. Waegell, P. K.  Aravind, {\it J. Phys. A: Math. Theor.} {\bf 45} 405301 (2012). DOI: 10.1088/1751-8113/45/40/405301
\bibitem{R17} J. Bub, {\it Found. Phys.} {\bf 26} 787 (1996). DOI: 10.1007/BF02058633
\bibitem{R18} M. Waegell, P. K. Aravind, {\it Found. Phys. } {\bf 41} 1786 (2011). DOI: 10.1007/s10701-011-9578-8
\bibitem{R19} M. Waegell, P. K. Aravind, {\it J. Phys. A: Math. Theor.} {\bf 44} 505303 (2011). DOI: 10.1088/1751-8113/44/50/505303
\bibitem{R20} G. Ga$\tilde{\mbox{n}}$as, S. Etcheverry, E. S. G$\acute{\mbox{o}}$mez, C. Saavedra, G. B. Xavier, G. Lima, A. Cabello, {\it Phys. Rev.} A {\bf 90} 012119 (2014). DOI: http://dx.doi.org/10.1103/PhysRevA.90.012119
\end{thebibliography}
\end{document}